\documentstyle[preprint,aps,version2]{revtex}
\begin{document}

\bibliographystyle{unsrt}
\def\Journal#1#2#3#4{{#1} {\bf #2}, #3 (#4)}
\def\NCA{ Nuovo Cimento}
\def\NIM{ Nucl. Instrum. Methods}
\def\NIMA{{ Nucl. Instrum. Methods} A}
\def\NP{{ Nucl. Phys.} }
\def\PLB{{ Phys. Lett.}  B}
\def\PRL{ Phys. Rev. Lett.}
\def\PRD{{ Phys. Rev.} D}
\def\ZPC{{Z. Phys.} C}
\def\ZPA{{Z. Phys.} A}
\def\st{\scriptstyle}
\def\sst{\scriptscriptstyle}
\def\mco{\multicolumn}
\def\epp{\epsilon^{\prime}}
\def\vep{\varepsilon}
\def\ra{\rightarrow}
\def\ppg{\pi^+\pi^-\gamma}
\def\vp{{\bf p}}
\def\ko{K^0}
\def\kb{\bar{K^0}}
\def\al{\alpha}
\def\ab{\bar{\alpha}}
\def\be{\begin{equation}}
\def\ee{\end{equation}}
\def\bea{\begin{eqnarray}}
\def\eea{\end{eqnarray}}
\def\CPbar{\hbox{{\rm CP}\hskip-1.80em{/}}}
\def\ua{\uparrow}
\def\da{\downarrow}

\begin{titlepage} 

\begin{title}
The hard scattering amplitude for the higher helicity components
of the pion form factor\thanks{Work partially supported 
by the National Natural Science Foundation of China 
and the National Education Committee of China}
\end{title}

\author{
Fu-Guang Cao$^{a,b,c,}$\thanks{E-mail address: caofg@itp.ac.cn.},
Jun Cao$^b$, Tao Huang$^{a,b}$, and Bo-Qiang Ma$^{a,b}$
}

\begin{instit}
$^a$ CCAST (World Laboratory), P.O. Box 8730, Beijing 100080, P.~R.~China
\end{instit}
\begin{instit}
$^b$ Institute of High Energy Physics, Academia Sinica,
P.O. Box 918, Beijing 100039, P.~R.~ China
\end{instit}
\begin{instit}
$^c$ Institute of Theoretical Physics, Academia Sinica,
P.O. Box 2735, Beijing 100080, P.~R.~China\thanks{Mailing address.}
\end{instit}

\begin{abstract}
For obtaining the spin space wave function of the pion meson 
in the light-cone formalism from the naive quark model,
it is necessary to take into account Wigner rotation.
Consequently there are higher helicity
($\lambda_1+\lambda_2=\pm 1$) components in the light-cone spin space
wave function of the pion besides the usual helicity
($\lambda_1+\lambda_2=0$) components.
For the pion electromagnetic form factor,
we calculate the hard-scattering amplitude for the higher helicity components
in the light-cone perturbation theory.
It is found that
the hard-scattering amplitude for the higher helicity components
is of order $1/Q^4$, which 
is vanishingly small compared to 
that of the ordinary helicity component
at very high $Q^2$ but
should be considered in the $Q^2$ region where 
experimental data are available.

\bigskip
\noindent
PACS number(s): 12.38.Bx, 12.39.Ki, 13.40.Gp, 14.40.Aq
\end{abstract}

\end{titlepage}

\section{Introduction}

There have been a lot of discussions about whether perturbative QCD (PQCD)
is applicable to exclusive processes at currently available experimental
energies [1-12]. In the example of the pion and proton form factors,
Isgur and Llewellyn Smith \cite{Isgur} noticed
that in the energy region of a few
GeV the main contributions come from the end-point region
$x\rightarrow 0,1$ ($x$ is the fractional momentum carried by the parton)
where the running couple constant $\alpha_s$ becomes large.
Thereby perturbation expansion might be illegal.
Recently this problem has been attacked and it 
is suggested that PQCD might be still applicable to the exclusive processes
at currently experimental accessible regime of momentum transfer
($Q^2 \sim $ a few GeV$^2$) by using some techniques to cure
the end-point problem [5-10]. 
Huang and Shen \cite{Huang} pointed out that the applicability of PQCD to
the hadronic form factors is questionable only as momentum transfers
being $Q^2\le 4$GeV$^2$ by reanalyzing the contributions from the end-point
region for the pion form factor.
Li and Sterman \cite{LiPion} proposed a modified perturbation expression for
the pion form factor by taking into account the customarily neglected
partonic transverse momentum as well as Sudakov correction.
They obtain a similar conclusion as \cite{Huang}: PQCD begins to be
self-consistent at about $Q\sim 20\Lambda_{QCD}$.
More recently, Ji, Pang, and Szczepaniak \cite{Ji95}
pointed out that the usual
factorization perturbation expression for the pion form factor is
derived from the light-cone time-order perturbative expansion,
and the natural variable to make a separation of perturbative
contributions from contributions intrinsic to the bound-state
wave function itself is the light-cone energy rather than the gluon
virtuality of the hard scattering amplitude $T_H$.
They find that the ``legal" PQCD contribution
defined by the light-cone energy cut becomes self-consistent at even
much smaller $Q^2$ region as compared to that defined by the gluon
four-momentum square cut.

Nevertheless, we notice that although most of the recently 
calculations [5-12] 
show that perturbative QCD is self-consistent and applicable 
to the exclusive processes
at currently experimental accessible energy regions,
the numerical predictions for the pion form factor are much smaller
than the experimental data.
There are two possible explanations:
one is that the non-perturbative contributions will dominate in this region;
the other is that the non-leading order contributions
in perturbative expansions may be also important in this region.
To make choice between the two possible explanations one needs
to analyse the non-leading contributions which
come from higher-twist effect,
higher order in $\alpha_s$, and higher Fock states {\it etc.}
Field, Gupta, Otto, and Chang \cite{Field}
pointed out that the contribution from the next-leading order
in $\alpha_s$ is about $20\%\sim 30\%$ to the perturbative pion form factor.
Employing the
modified factorization expression for the pion form factor proposed
by Li and Sterman \cite{LiPion},
Refs. \cite{Jacob,Caopion2} considered the transverse momentum
effect in the wave function and found that the transverse momentum
in the wave function play the role to suppress perturbative prediction.
Thus it is necessary to calculate the other non-leading contributions
such as that from higher twist effect and higher Fock states.

One of the other sources which may provide non-leading 
perturbative contribution
is the higher helicity components in the
light-cone wave function \cite{HMS,MaZPA,MaJG}. 
The effects from higher helicity components (or Wigner rotation effect) 
have been investigated in the description of pion properties at high
energies \cite{MaJG,Wang} as well as at low energies [15,18-21]
and the same effect has been also
applied to explain the ``proton spin crisis" \cite{crisis,Ma96}.
However, the calculations for
the contributions coming from higher helicity components
to the pion form factor in the high energy region
are conflicting in literatures \cite{MaJG,Wang}.
Ma and Huang \cite{MaJG} pointed out that
the higher helicity components provide a large {\sl enhancement}
for the perturbation prediction of the pion form factor and
thus may provide the other fraction 
which is needed to
fit the experimental data around $Q^2 \ge 2$~GeV$^2$.
More recently, Wang and Kisslinger \cite{Wang}
also analysed this effect based on the modified perturbative
approach. In their approach 
this effect gives a large {\sl suppression} for the pion form factor
as compared to the prediction obtained in the original hard-scattering
model in the $Q^2$ domain where experimental data are available.
Thereby they concluded that non-perturbative contributions 
dominate in this region.
Refs.~\cite{MaJG} and \cite{Wang} gave very different conclusions
concerning the question whether the perturbative QCD contributions
dominate or not in the available experimental energy region.
We point out that the conflict between the above works is 
due to the difference between 
the hard-scattering amplitudes for the higher helicity components
adopted in Refs.~\cite{MaJG} and \cite{Wang}.
It is assumed in Ref.~\cite{MaJG}
that the hard-scattering amplitude for the
higher helicity components is the same as that for the ordinary helicity
component,
\bea
T_H^{(\lambda_1+\lambda_2=\pm 1)}=T_H^{(\lambda_1+\lambda_2=0)}
=\frac{4g^2C_F}{x_2y_2 Q^2}.
\label{hsaMH}
\eea
But the hard-scattering amplitude employed in Ref.~\cite{Wang} is
\bea
T_H^{(\lambda_1+\lambda_2=\pm 1)}=-T_H^{(\lambda_1+\lambda_2=0)}
=-\frac{4g^2C_F}{x_2y_2Q^2+({\bf k}_\perp-{\bf l}_\perp)^2}
\stackrel{Q^2\rightarrow\infty} \approx -\frac{4g^2C_F}{x_2y_2 Q^2}.
\label{hsaKW}
\eea
It can be seen that
the asymptotic ($Q^2\rightarrow \infty$) behaviors of 
Eqs. (\ref{hsaMH}) and (\ref{hsaKW}) are with {\sl opposite} signs. 
That is the reason that Refs.~\cite{MaJG} and \cite{Wang}
gave {\sl opposite} conclusions concerning the PQCD contributions
from the higher helicity components in the experimental available
$Q^2$ region.

The purpose of this paper is to analyze the effect from 
the higher-helicity 
components of the pion wave function 
in the light-cone perturbative QCD and address the conflict between
Refs.~\cite{MaJG} and \cite{Wang}.
We first review and analyze the spin structure for the pion
light-cone wave function and the necessity to take into account the higher 
helicity
components in Sec.~II. Then in 
Sec.~III we calculate the
hard-scattering amplitude for the higher helicity components
of the pion form factor. 
We first explicitly show that the hard-scattering amplitude
for the higher helicity components vanishes in the leading order $O(1/Q^2)$
As the parton intrinsic transverse momentum is taken into account,
it is found that the asymptotic behavior of the hard-scattering amplitude
for the higher helicity components is of order $1/Q^4$
which differs from either Eq.~(\ref{hsaMH}) or Eq.~(\ref{hsaKW}).
We conclude that the higher helicity components, though
provide vanishingly small contributions 
to the perturbative pion form factor
in the asymptotic limit $Q^2 \to \infty $,
they should be considered in the 
available experimental energy region since they are
next-to-leading order contributions.
Sec.~IV is served as a summary.

\section{
The light-cone wave function of the pion and its higher helicity 
components}

The light-cone (LC) formalism \cite{LCF} provides a
convenient framework for the relativistic description of hadrons
in terms of quark and gluon degrees of freedom,
and the application of PQCD to
exclusive processes has mainly been developed in
this formalism (light-cone PQCD) \cite{Brodsky,BrodskyPQCD,BHL}.
The essential feature of light-cone PQCD 
application to exclusive processes is
that the amplitudes for these processes can be written as a
convolution of hadron light-cone wave functions (or quark distribution
amplitudes)
for every hadron involved in the process with a hard-scattering
amplitude $T_H$. Thus light-cone wave function is
an essential part: It determines the distributions of the quark and
gluons entering the short distance sub-processes and provides the link
between the long-distance non-perturbative and short distance
perturbative physics.
In principle, light-cone wave function can be computed from rigorous 
light-cone QCD.
Unfortunately this task is very complex and difficult, and
there is no exact solution up to now.
More practical and more convenient way is 
to connect light-cone wave function with
the instant-form wave function which can be obtained by solving
the Bethe-Salpeter equation with some approximations \cite{BHL}.
The connection for the spin space wave functions between the two formalisms 
are accomplished \cite{HMS,MaZPA,MaJG} by the use of Wigner 
rotation \cite{Wigner}.
The connection for the momentum space wave functions become possible
with the help of some ansatz such as the Brodsky-Huang-Lepage
prescription \cite{BHL}.

It should be emphasized that in order to connect the spin structures in
two different frames correctly, 
it is necessary to consider Wigner rotation 
effect.
As it is known, spin is essentially a relativistic notion associated with
the space-time symmetry of Poincar\'{e}.
The conventional 3-vector spin
{\bf s} of a moving particle with finite mass $m$ and 4-momentum $p_\mu$
can be defined by transforming its Pauli-Lub\'{a}nski 4-vector
$\omega_\mu=1/2 J^{\rho\sigma}P^\nu\epsilon_{\nu\rho\sigma\mu}$
to its rest frame via a non-rotation Lorentz boost $L(p)$ which
satisfies $L(p)p=(m,{\bf 0})$, by $(0,{\bf s})=L(p)\omega/m$.
Under an arbitrary Lorentz transformation, 
a particle state with spin ${\bf s}$
and 4-momentum $p_\mu$ will transform to the state with spin ${\bf s}'$ and
4-momentum $p'_\mu$,
\bea
{\bf s}'=R_\omega({\bf \Lambda},p){\bf s}, ~~~~ p'={\bf \Lambda}p,
\eea
where $R_\omega({\bf \Lambda},p)=L(p'){\bf \Lambda} L^{-1}(p)$
is a pure rotation known as Wigner rotation.
When a composite system is transformed from one frame to another one,
the spin of each constituent will undergo a Wigner rotation.
These spin rotations
are not necessarily the same since the constituents have different internal
motion. In consequence, the sum of the constituent's spin is not Lorentz
invariant. Hence, although the pion has only $\lambda_1+\lambda_2=0$
spin components in the rest frame of the pion, it may have
$\lambda_1+\lambda_2=\pm 1$ spin components in the infinite-momentum
frame (light-cone formalism)\footnote{Notice that the instant-form dynamics
in the infinite-momentum frame is equivalent to light-front dynamics
in an ordinary frame.}, where $\lambda_1$ and $\lambda_2$ are the
quark and anti-quark helicities respectively.
One advantage of light-cone dynamics is that Wigner rotation relating
spin states in different frames is unity under a kinematic
Lorentz transformation.

To obtain light-cone spin space wave function of the pion one can transform
the ordinary instant-form SU(6) quark model spin space wave function
of the pion into light-cone dynamics \cite{HMS,MaZPA,MaJG}.
In the pion rest frame $({\bf q}_1+{\bf q}_2=0)$, 
the instant-form spin space
wave function of the pion is
\bea
\chi_T=(\chi_1^\ua \chi_2^\da-\chi_2^\ua \chi_1^\da)/\sqrt{2},
\label{spinor-Inst}
\eea
in which $\chi_i^{\ua,\da}$ are the two-component Pauli spinors and
$q_1^\mu=(q^0,{\bf q})$, $q_2^\mu=(q^0,-{\bf q})$ are 4-momenta for
the two quarks respectively with $q^0=(m^2+{\bf q}^2)^{1/2}$.
The instant-form spin states $|J,s\rangle_T$ and the light-cone form
spin states  $|J,s\rangle_F$ are related by a Wigner
rotation $U^J$ [15-21], 
\bea
|J,\lambda\rangle_F=\sum_s U_{s\lambda}^J |J,s\rangle_T.
\label{Wigner-rotation}
\eea
This rotation is called as Melosh rotation \cite{Melosh}
for spin-1/2 particles.
Applying transformation Eq.~(\ref{Wigner-rotation})
on the both sides of Eq.~(\ref{spinor-Inst}) one can obtain
the spin space wave function of the pion in the infinite-momentum frame.
Transforming for the left side ({\it i.e.}, the pion)
is simple since Wigner
rotation is unity. For the right side ({\it i.e.}, the two spin-1/2
partons),
each particle instant-form and light-cone form spin states are related
by the Melosh transformation,
\bea
\chi^\ua(T)&=&w[(q^++m)\chi^\ua(F)-q^R\chi^\da(F)];\nonumber\\
\chi^\da(T)&=&w[(q^++m)\chi^\da(F)+q^L\chi^\ua(F)],
\label{Melosh-rotation}
\eea
where $w=[2q^+(q^0+m)]^{-1/2}$, $q^{R,L}=q^1\pm i q^2$, and
$q^+=q^0+q^3$. Then the light-cone spin space wave function of the pion reads
\bea
\chi_F(x, {\bf k}_\perp)=\sum_{\lambda_1,\lambda_2}
C_0^F(x,{\bf k}_\perp, \lambda_1,\lambda_2)
\chi_1^{\lambda_1}(F)\chi_2^{\lambda_2}(F).
\label{spinor-LC}
\eea
When expressed
in terms of the equal-time momentum $q^{\mu}=(q^{0},{\bf q})$, the
spin component coefficients $C_0^F$ have the forms,
\bea
    C^{F}_{0}(x,q,\uparrow,\downarrow)&=&
\displaystyle{
w_{1}w_{2}[(q_{1}^{+}+m)(q_{2}^{+}+m)-{\bf q}_{\perp}^{2}]/\sqrt{2}
};\nonumber\\
    C^{F}_{0}(x,q,\downarrow,\uparrow)&=&-
\displaystyle{
w_{1}w_{2}[(q_{1}^{+}+m)(q_{2}^{+}+m)-{\bf q}_{\perp}^{2}]/\sqrt{2}
};\nonumber\\
    C^{F}_{0}(x,q,\uparrow,\uparrow)&=&
\displaystyle{
w_{1}w_{2}[(q_{1}^{+}+m)q_{2}^{L}-(q_{2}^{+}+m)q_{1}^{L}]/\sqrt{2}
};\label{coefficient1}\\
    C^{F}_{0}(x,q,\downarrow,\downarrow)&=&
\displaystyle{
w_{1}w_{2}[(q_{1}^{+}+m)q_{2}^{R}-(q_{2}^{+}+m)q_{1}^{R}]/\sqrt{2}
}.\nonumber
\eea
The equal-time momentum ${\bf q}=(q^{3},{\bf q}_{\perp})$
and the light-cone momentum $\underline{k}=(x,{\bf k}_{\perp})$
can be connected according to 
the Brodsky-Huang-Lepage prescription \cite{BHL}
which is obtained by  equating the off-shell propagators in the two frames,
\be
\begin{array}{ccc}
      xM & \leftrightarrow & (q^{0}+q^{3});\\
      {\bf k}_{\perp} & \leftrightarrow & {\bf q}_{\perp},
\end{array}
\label{relation}
\ee
in which $M$ is defined as
\begin{equation}
       M^{2}=\frac{{\bf k}_{\perp}^{2}+m^{2}}{x(1-x)}.
\label{Mdefination}
\end{equation}
From (\ref{relation}) we have
\begin{equation}
\frac{{\bf k}_{\perp}^{2}+m^{2}}{4x(1-x)}-m^{2}={\bf q}^{2}.
\end{equation}
From Eqs.~(\ref{coefficient1}), (\ref{relation}) and
(\ref{Mdefination}) the coefficients $C_0^F$ can be
expressed in the light-cone momentum $\underline{k}=(x,{\bf k}_{\perp})$,
\bea
C_0^F(x,{\bf k}_\perp,\ua,\da)&=&
\frac{m}{[2(m^2+{\bf k}_\perp^2)]^{1/2}};\nonumber\\
C_0^F(x,{\bf k}_\perp,\da,\ua)&=&
-\frac{m}{[2(m^2+{\bf k}_\perp^2)]^{1/2}}; \nonumber\\
C_0^F(x,{\bf k}_\perp,\ua,\ua)&=&
-\frac{(k_1-ik_2)}{[2(m^2+{\bf k}_\perp^2)]^{1/2}};
\label{coefficient2}\\
C_0^F(x,{\bf k}_\perp,\da,\da)&=&
-\frac{(k_1+ik_2)}{[2(m^2+{\bf k}_\perp^2)]^{1/2}}. \nonumber
\eea
$C_0^F$ satisfy the relation
\bea
\sum_{\lambda_1,\lambda_2}C_0^F(x,{\bf k}_\perp,\lambda_1,\lambda_2)
C_0^F(x,{\bf k}_\perp,\lambda_1,\lambda_2)=1.
\eea
It can be seen explicitly from Eqs.~(\ref{spinor-Inst}),
(\ref{spinor-LC}) and (\ref{coefficient2})
that the light-cone
spin space wave function of the pion $\chi_F$ has higher helicity
($\lambda_1+\lambda_2=\pm 1$) components besides the ordinary helicity
($\lambda_1+\lambda_2=0$) component, while the instant-form
spin space wave function of the pion, $\chi_T$
 has only the ordinary helicity component.
Notice that $\chi_F$ is also an eigen-state of the total spin operator
$(\hat{S}^{F})^{2}$ in the light-cone formalism \cite{MaJG}.

Now the light-cone wave function for the lowest valence state of the pion
can be expressed as \cite{MaJG}
\begin{eqnarray}
|\psi^{\pi}_{q\overline{q}}>=
\psi(x,{\bf k}_{\perp},\uparrow,\downarrow)|\uparrow\downarrow>
+\psi(x,{\bf k}_{\perp},\downarrow,\uparrow)|\downarrow\uparrow>
\nonumber \\
+\psi(x,{\bf k}_{\perp},\uparrow,\uparrow)|\uparrow\uparrow>
+\psi(x,{\bf k}_{\perp},\downarrow,\downarrow)|\downarrow\downarrow>,
\label{eq:pwf}
\end{eqnarray}
where
\begin{equation}
\psi(x,{\bf k}_{\perp},\lambda_{1},\lambda_{2})=
C^{F}_{0}(x,{\bf k}_{\perp},\lambda_{1},\lambda_{2})
\varphi(x,{\bf k}_{\perp}).
\end{equation}
Here $\varphi(x,{\bf k}_{\perp})$ is the momentum space wave function
in the light-cone formalism.

The above result means that the light-cone spin of a
composite particle is not directly the sum of its constituents'
light-cone spins but the sum of Wigner rotated light-cone spins of the
individual constituents.
A natural consequence is that in light-cone formalism a hadron's
helicity is not necessarily equal to the sum of the
quark's helicities, {\it i.e.}, $\lambda_{H}\neq \sum_{i} \lambda_{i}$.
This result is important for understanding the proton ``spin
puzzle'' \cite{crisis}. 
It has been shown \cite{Ma96} that the relativistic SU(6) quark model of the 
nucleon,
supplemented with Wigner rotation effect \cite{crisis} and 
the flavor asymmetry generated by the spin-spin interaction
of the valence spectator quarks, could reproduce the observed ratio
$F_2^n/F_2^p$ and the proton, neutron, and deuteron polarization asymmetries,
$A_1^p$, $A_1^n$, $A_1^d$.
If the 
intrinsic quark-antiquark pairs generated by the
non-perturbative meson-baryon fluctuations in the nucleon sea
are further taken into account, we could arrive at a
consistent framework \cite{Bro96} to understand a number of  
anomalies observed 
in the
proton's structure: the origin of polarized strange quarks implied by
the violation of the  Ellis-Jaffe sum rule;  the flavor asymmetry of the
nucleon sea implied by the violation of Gottfried sum rule;
and the conflict between two different measurements of strange
quark distributions.

\section{The hard scattering amplitude for the higher helicity components
in the pion form factor}

The pion electromagnetic form factor can be expressed by the
Drell-Yan-West formula \cite{Drell},
\be
F(Q^2)=\sum_{n,\lambda_i}\sum_j e_j \int [dx][d^2{\bf k}_\perp]
\psi^*_n(x_i,{\bf k}_{\perp,i},\lambda_i)
\psi_n(x_i,{\bf k}'_{\perp,i},\lambda_i),
\label{Drell-Yan}
\ee
where ${\bf k}'_\perp={\bf k}_\perp-x_i{\bf q}_\perp +{\bf q}_\perp$
for the struck quark,
${\bf k}'_\perp={\bf k}_\perp-x_i{\bf q}_\perp $
for the spectator quarks, and $e_i$ is the electric charge
of the struck quark. At higher momentum transfer, the pion form factor
in the leading order can be given by \cite{LiPion,Ji95,MaJG}
\bea
F_\pi(Q^2)&=&\int [dx][dy]\int[d^2{\bf k}_\perp]
[d^2{\bf l}_\perp]\psi^{((1-x)Q)}(x,{\bf k}_\perp,\lambda_i)
T_H(x,y,{\bf q}_\perp,{\bf k}_\perp,{\bf l}_\perp)
{\psi^*}^{((1-y)Q)}(y,{\bf l}_\perp,\lambda_i) \nonumber\\
&=&\int [dx][dy]\int[d^2{\bf k}_\perp]
[d^2{\bf l}_\perp]\varphi^{((1-x)Q)}(x,{\bf k}_\perp)
[{\cal W}_1 T_H^{(\lambda_1+\lambda_2=0)}(x,y,{\bf q}_\perp,{\bf k}_\perp,
{\bf l}_\perp) \nonumber\\
& &~~~~~~~~~~~~~~~~~~~~~~~~~~~~~~~
+{\cal W}_2 T_H^{(\lambda_1+\lambda_2=\pm 1)}
(x,y,{\bf q}_\perp,{\bf k}_\perp,{\bf l}_\perp)]
{\varphi^*}^{((1-y)Q)}(y,{\bf l}_\perp)
\label{Fpion}
\eea
where $[dx]=dx\delta(1-x_1-x_2)$,
$[d^2{\bf k}_\perp]={d^2{\bf k}_\perp}/{(16\pi^3)}$,
$\varphi^{((1-x)Q)}(x,{\bf k}_\perp)$ is the light-cone momentum space
wave function of the valence Fock state with a cut-off
${\bf k}_\perp^2=(1-x)Q$, $T_H$ are the hard-scattering amplitudes
which can be calculated from the 
time-ordered diagrams in light-cone PQCD,
and ${\cal W}_1$ and ${\cal W}_2$ are the factors from Wigner rotation,
\bea
{\cal W}_1&=&m/\left[(m^2+{\bf k}_\perp^2)(m^2+{\bf l}_\perp^2)
\right]^{1/2};\nonumber \\
{\cal W}_2&=&{\bf k}_\perp\cdot{\bf l}_\perp
/\left[(m^2+{\bf k}_\perp^2)(m^2+{\bf l}_\perp^2)\right]^{1/2}.
\eea
In the derivation for Eq.~(\ref{Fpion}) we have applied the relations
\bea
T_H^*(\da\ua\rightarrow\da\ua)=T_H(\ua\da\rightarrow\ua\da),~~~~
T_H^*(\da\da\rightarrow\da\da)=T_H(\ua\ua\rightarrow\ua\ua).
\label{THrelation}
\eea
After summing over all helicities, only the real part
of each hard-scattering amplitude survives. Thereby there are
only two independent hard-scattering amplitudes:
\bea
T_H^{(\lambda_1+\lambda_2=0)}&=& \frac{1}{2}
\left[T_H(\ua\da\rightarrow\ua\da)+T_H(\da\ua\rightarrow\da\ua)
\right];\nonumber\\
T_H^{(\lambda_1+\lambda_2=\pm 1)}&=&\frac{1}{2}
\left[T_H(\ua\ua\rightarrow\ua\ua)+T_H(\da\da\rightarrow\da\da)\right].
\label{THdefination}
\eea
As ${\bf k}_\perp={\bf l}_\perp=0$, Wigner rotation factors
${\cal W}_1=1$, ${\cal W}_2=0$, and Eq.~(\ref{Fpion})
reduces to the ordinary perturbation expression for the pion form factor.
In more general situation, 
there is also contribution from the higher helicity components 
$T_H^{(\lambda_1+\lambda_2=\pm 1)}$ besides 
the hard-scattering amplitude $T_H^{(\lambda_1+\lambda_2=0)}$
from the ordinary helicity component of the pion.
Notice that quark helicity is conserved at each vertex in $T_H$ in the
limit of vanishing quark mass, since both photon and gluon are vector
particles \cite{Brodsky,BL2}. 
Hence there is no hard-scattering amplitude with
quark and antiquark helicities being changed.
$T_H^{(\lambda_1+\lambda_2=0)}$ has been
calculated 
in cases when the intrinsic transverse momenta are neglected
(see for example \cite{Brodsky,BrodskyPQCD,BHL}) and taken into account
\cite{Ji95}.
The purpose of this paper is to
calculate
$T_H^{(\lambda_1+\lambda_2=\pm 1)}$, {\it i.e.}, the contribution
from the higher helicity components of the pion light-cone 
wave function.

In the light-cone perturbative QCD, 
there are six time-order diagrams as shown in Fig.~1
which contribute to $T_H(\ua\ua\rightarrow\ua\ua)$ and
$T_H(\da\da\rightarrow\da\da)$.
The calculation rules for the light-cone PQCD can be found
in literatures \cite{Brodsky,BrodskyPQCD,BHL}.
First, we neglect the intrinsic transverse momenta
${\bf k}_\perp$ and ${\bf l}_\perp$. The contribution of diagram (a)
can be written as,
\bea
T_H^{(a)}={\rm Tr} \frac{1}{D_{11}}\frac{1}{D_{12}}
	  \frac{\theta(y_1-x_1)}{y_1-x_1} +{\rm Inst.},
\label{THaLO}
\eea
where $D_{11}$ and $D_{12}$ are the ``energy denominators",
\bea
D_{11}=-\frac{y_1x_2^2}{x_1(y_1-x_1)}{\bf q}_\perp^2,~~~~
D_{12}=-\frac{x_2}{x_1}{\bf q}_\perp^2,
\eea
and ${\rm Tr}$ is the sum of some spinors and $\gamma$-matrix in 
light-cone PQCD,
\bea
{\rm Tr}=\frac{\bar{u}_\ua(y_1,y_1{\bf q}_\perp)}{\sqrt{y_1}}ig\gamma^\mu
       \frac{u_\ua(x_1,{\bf q}_\perp)}{\sqrt{x_1}}
        d_{\mu\nu}\frac{\bar{v}_\da(x_2,{\bf o}_\perp)}
        {\sqrt{x_2}}ig\gamma^\nu
	\frac{v_\ua(y_2,y_2{\bf q}_\perp)}{\sqrt{y_2}}.
\eea
By using Eqs.~(\ref{THrelation}) and (\ref{THdefination}),
we need to calculate only the real part of ${\rm Tr}$ which reads,
\bea
{\rm RTr}=-g^2\frac{2x_2(x_1y_2+y_1x_2)}{x_1(y_1-x_1)^2}{\bf q}_\perp^2.
\label{RTrLO}
\eea
The ``${\rm Inst.}$" part in Eq.~(\ref{THaLO}) represents the contribution 
from instantaneous diagram which is one feature of light-cone PQCD,
\bea
{\rm Inst.}=g^2\frac{4x_1\theta(y_1-x_1)}{x_2(y_1-x_1)^2{\bf q}_\perp^2}.
\eea
Then the contribution from diagram (a) reads 
\bea
T_H^{(a)}=g^2\frac{2x_1}{x_2^2y_1{\bf q}_\perp^2}
\frac{\theta(y_1-x_1)}{y_1-x_1}.
\eea
It is known that
the contribution from each diagrams, for example 
$T_H^{(a)}$, is itself not
gauge-invariant, but the gauge-invariance will be satisfied
when 
summing over all time-order diagrams (a)-(f).
The contributions from the other diagrams can be calculated
in a similar way.
Observing that the term ``${\rm Tr}$" is the same for the diagrams
(a), (b) and (c), and employing the following relations for the
``energy denominators",
\bea
D_{22}=D_{12}, &~~~~& D_{31}=D_{11},\nonumber \\
D_{11}=D_{32}+D_{12}, &~~~~&
D_{21}=-D_{32}=\frac{x_2y_2}{y_1-x_1}{\bf q}_\perp^2,
\eea
we can sum over 
the contributions from diagrams (a), (b) and (c),
\bea
T_H^{(a+b+c)}&=&{\rm Tr}\frac{1}{D_{12}}\frac{1}{D_{21}}\frac{1}{x_1-y_1}
-g^2\frac{1}{D_{12}}\frac{4}{(x_1-y_1)^2} \nonumber\\
&=&\frac{2g^2}{x_2y_2{\bf q}_\perp^2}\frac{1}{x_1-y_1}.
\label{TH-abc}
\eea
We point out that under transformation ($x \leftrightarrow y$)
there is symmetry for the six diagrams,
\bea
{\rm Diagrams}~(a,b,c) \Longleftrightarrow {\rm Diagrams}~(d,e,f)
~~~~{\rm under}~~(x\leftrightarrow y).
\eea
Thus the contributions form diagrams (d),
(e), and (f) are,
\bea
T_H^{(d+e+f)}=\frac{2g^2}{x_2y_2{\bf q}_\perp^2}\frac{1}{y_1-x_1}.
\label{TH-def}
\eea
From Eqs.~(\ref{TH-abc}) and (\ref{TH-def})
we can obtain the hard-scattering amplitude
for the higher helicity components in the approximation neglecting
parton intrinsic transverse momenta,
\bea
T_H^{(\lambda_1+\lambda_2=\pm 1)}(x,y,{\bf q}_\perp)=
T_H^{(a+b+c)}+T_H^{(d+e+f)}=0.
\label{TH1nkl}
\eea
Eq.~(\ref{TH1nkl}) shows that there is no contribution from the 
higher helicity components for the pion form factor
in the leading order $O(1/Q^2)$
(or the intrinsic transverse momenta being neglected);
which is in agreement with the early result obtained by Brodsky
and Lepage \cite{Brodsky,BL2}.

Now we take into account the parton intrinsic transverse momenta
${\bf k}_\perp$ and ${\bf l}_\perp$. Then ``${\rm Tr}$" means
\bea
{\rm Tr}=\frac{\bar{u}_\ua(y_1,y_1{\bf q}_\perp
+{\bf l}_\perp)}{\sqrt{y_1}}ig\gamma^\mu
       \frac{u_\ua(x_1,{\bf q}_\perp+{\bf k}_\perp)}{\sqrt{x_1}}
        d_{\mu\nu}\frac{\bar{v}_\da(x_2,-{\bf k}_\perp)}
        {\sqrt{x_2}}ig\gamma^\nu
	\frac{v_\ua(y_2,y_2{\bf q}_\perp-{\bf l}_\perp)}{\sqrt{y_2}}
\eea
and
\bea
{\rm RTr}=\frac{[y_1(x_2{\bf q}_\perp+{\bf k}_\perp)-x_1{\bf l}_\perp]
\cdot [y_2(x_2{\bf q}_\perp+{\bf k}_\perp)-x_2{\bf l}_\perp]}
{x_1x_2y_1y_2(x_1-y_1)^2}[2(x_1y_2+y_1x_2)].
\label{RTr1}
\eea
The ``energy denominators" are
\bea
D_{11}=-\frac{(x_2{\bf q}_\perp+{\bf k}_\perp)^2}{x_1x_2}
       -\frac{[y_2(x_2{\bf q}_\perp+{\bf k}_\perp)-x_2{\bf l}_\perp]^2}
         {x_2y_2(y_1-x_1)}, &~~~~&
D_{12}=-\frac{(x_2{\bf q}_\perp+{\bf k}_\perp)^2}{x_1x_2};\nonumber \\
D_{21}=-\frac{{\bf l}_\perp^2}{y_1y_2}
       +\frac{[y_2(x_2{\bf q}_\perp+{\bf k}_\perp)-x_2{\bf l}_\perp]^2}
         {x_2y_2(y_1-x_1)},&~~~~&
D_{22}=D_{12};\\\label{Drelation1}
D_{32}=-\frac{{\bf k}_\perp^2}{x_1x_2}
       -\frac{[y_2(x_2{\bf q}_\perp+{\bf k}_\perp)-x_2{\bf l}_\perp]^2}
         {x_2y_2(y_1-x_1)},&~~~~&
D_{31}=D_{11}. \nonumber
\eea
Using the symmetry
\bea
{\rm Diagrams}~(a,b,c)\Longleftrightarrow {\rm Diagrams}~(d,e,f)
~~~~ {\displaystyle{{\rm under}}}  ~~
\left\{
\begin{array}{lcr}
x &\leftrightarrow& y \\
{\bf k}_\perp &\leftrightarrow& -{\bf l}_\perp
\end{array}
\right\}
\label{symmetry}
\eea
we get,
\bea
T_H^{(\lambda_1+\lambda_2=\pm 1)}(x,y,{\bf q}_\perp,{\bf k}_\perp,
{\bf l}_\perp)&=&
{\rm RTr }\left[\left(\frac{1}{D_{11}D_{12}}+\frac{1}{D_{31}D_{32}}\right)
\frac{\theta(y_1-x_1)}{y_1-x_1}+\frac{1}{D_{21}D_{22}}
\frac{\theta(x_1-y_1)}{x_1-y_1}\right]\nonumber\\
&&+\frac{4}{D_{12}(y_1-x_1)^2}+\left\{
\begin{array}{lcr}
x&\leftrightarrow&y\\
{\bf k}_\perp&\leftrightarrow&-{\bf l}_\perp
\end{array}
\right\}.
\label{TH1}
\eea
In the above calculation we have neglected the quark masses
since it is ``current quark masses" that should appear in perturbative 
calculation. The pionic mass can also be neglected in PQCD calculation.

To simplify Eq.~(\ref{TH1}), we adopt the following two prescriptions:
1) It is pointed out in Ref.~\cite{Ji95} that
as one concerns with the effect from intrinsic transverse momenta
the terms proportional to the ``bound energies" of the pions in the initial
and final states {\it i.e.} $\sim {\bf k}_\perp^2/(x_1x_2)$
and $\sim {\bf l}_\perp^2/(y_1y_2)$ can be ignored to avoid
the involvement of the higher Fock states 
contributions\footnote{As the transverse momenta 
${\bf k}_\perp$ and ${\bf l}_\perp$
are included, it is necessary to take into account the contributions 
from higher Fock states to satisfy the gauge-invariance, since
the covariant derivative $D_\mu=\partial_\mu+igA_\mu$ makes both 
transverse momenta ${\bf k}_\perp$, ${\bf l}_\perp$ and
the transverse gauge degree $g{\bf A}_\perp$ be of the same order \cite{Ji95}.}. 
Neglecting these terms in the ``energy denominators", we have,
\bea
T_H^{(\lambda_1+\lambda_2=\pm 1)}(x,y,{\bf q}_\perp,{\bf k}_\perp,
{\bf l}_\perp)&=&
\frac{[y_2(x_2{\bf q}_\perp+{\bf k}_\perp)-x_2{\bf l}_\perp]
\cdot [y_1(x_2{\bf q}_\perp+{\bf k}_\perp)+x_1{\bf l}_\perp]}
{(x_2{\bf q}_\perp^2+{\bf k}_\perp)^2
[y_2(x_2{\bf q}_\perp+{\bf k}_\perp)-x_2{\bf l}_\perp]^2}\nonumber \\
&& \times \frac{2g^2x_2}{y_1(x_1-y_1)}+\left\{
\begin{array}{lcc}
x&\leftrightarrow&y\\
{\bf k}_\perp&\leftrightarrow&-{\bf l}_\perp
\end{array}
\right\}.
\label{TH1-approximation1}
\eea
2) Notice that in the factorization expression for the pion form factor
Eq.~(\ref{Fpion}), we have ${\bf k}_\perp^2\ll{\bf q}_\perp^2$ and
${\bf l}_\perp^2\ll{\bf q}_\perp^2$. 
Hence when calculating to the next-to-leading order in $1/Q$ for $T_H$,
we can neglect the terms such as
${\bf k}_\perp^2/{\bf q}_\perp^2$, ${\bf k}_\perp^2/{\bf q}_\perp^2$
and $({\bf k}_\perp\cdot{\bf l}_\perp)/{\bf q}_\perp^2$ in the
both the ``energy denominators" and ``${\rm RTr}$". Then we get
\bea
T_H^{(\lambda_1+\lambda_2=\pm 1)}(x,y,{\bf q}_\perp,{\bf k}_\perp,
{\bf l}_\perp)&=&
\frac{[y_1y_2(x_2{\bf q}_\perp^2+2{\bf q}_\perp\cdot{\bf k}_\perp)
+(x_1-y_1){\bf q}_\perp\cdot{\bf l}_\perp]}
{(x_2{\bf q}_\perp^2+2{\bf q}\cdot {\bf k})
[y_2(x_2{\bf q}_\perp^2+2{\bf q}\cdot {\bf k})
-2x_2{\bf q}_\perp\cdot {\bf l}_\perp]}\nonumber \\
&&\times \frac{2g^2}{y_1y_2(x_1-y_1)}+\left\{
\begin{array}{lcc}
x&\leftrightarrow&y\\
{\bf k}_\perp&\leftrightarrow&-{\bf l}_\perp
\end{array}
\right\}.
\label{TH1-approximation2}
\eea

As the intrinsic transverse
momenta ${\bf k}_\perp$ and ${\bf l}_\perp$ are neglected (or in the 
asymptotic limit $Q^2 \rightarrow \infty$),
Eqs.~(\ref{TH1}), (\ref{TH1-approximation1}), and (\ref{TH1-approximation2})
reduce to Eq.~(\ref{TH1nkl}), {\sl i.e.}, the hard-scattering amplitude for
the higher helicity components goes to zero.
It can be found from Eqs. (\ref{TH1-approximation1})
and (\ref{TH1-approximation2}) that the leading contribution of
the hard-scattering amplitude for the higher helicity components 
is of order $1/Q^4$ which is next-to-leading contribution
compared to the contribution coming from the ordinary helicity component,
but it may give sizable contributions to the 
pion form factor in the intermediate energy region.
We also notice that (\ref{TH1-approximation1}) and
Eq.~(\ref{TH1-approximation2})
differ to either Eq.~(\ref{hsaMH}) or
Eq.~(\ref{hsaKW}), 
hence the calculations in neither Ref.~\cite{MaJG} nor Ref.~\cite{Wang}
is reliable.
It is necessary to re-consider the PQCD contributions
from the higher helicity components
based on proper hard-scattering amplitude derived
from theory at the energy scale where the 
current experiments are accessible.
The quantitative predictions depend on numerical calculation which
involves 6-dimensional integral with tedious technical details,
and will be given elsewhere.

\section{Summary}

The light-cone formalism provides a
convenient framework for the relativistic description of hadrons
in terms of quark and gluon degrees of freedom, and the application
of perturbative QCD to exclusive processes has mainly been developed in this 
formalism.
In order to obtain correct spin structure for the hadron wave function in
the light-cone formalism from the instant-form wave function, the 
relativistic effect due to Wigner rotation should be taken into account.
Consequently, in the light-cone formalism, 
there are higher helicity ($\lambda_1+\lambda_2=\pm 1$)
components in the spin space wave function besides the usual helicity
($\lambda_1+\lambda_2=0$) components .
We give the hard scattering
amplitude for the higher helicity components 
in the perturbative calculation for the pion form factor.
It is found that the hard-scattering amplitude
for the higher helicity components is of order $1/Q^4$,
which is vanishingly small compared to that of the ordinary helicity 
components at very high $Q^2$ but should be considered in the $Q^2$ region
where  experimental data are available.


\vskip 1cm
\section*{Figure caption}
\begin{description}
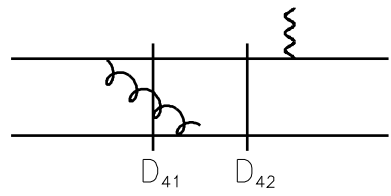
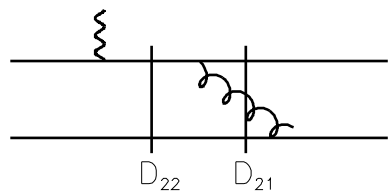
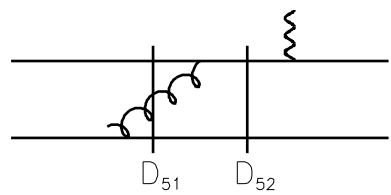
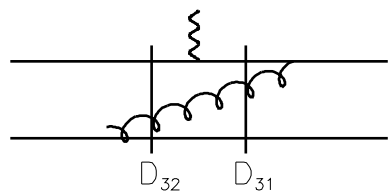
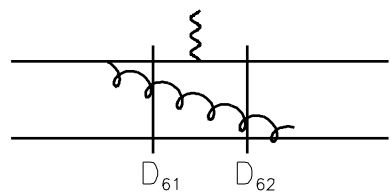
\item{Fig. 1.}
Leading order time-order diagrams contributing to the hard scattering
amplitude for the higher helicity $(\lambda_1+\lambda_2=\pm 1)$
components of the pion in the
perturbative calculation for the pion form factor,
where
$k_1=(x_1,{\bf k}_\perp)$,
$k_2=(x_2,-{\bf k}_\perp)$, $l_1=(y_1,y_1{\bf q}_\perp+{\bf l}_\perp)$,
and $l_2=(y_2,y_2{\bf q}_\perp-{\bf l}_\perp)$,
and the momenta are expressed in the light-cone variables $(+, \perp)$.
As usual the momentum of the pion in the initial state 
is taken to be  $P= (1, {\bf 0}_\perp)$ 
and the momentum of the photon is $q=(0,{\bf q}_\perp)$ with
$q^-={\bf q}_\perp^2$.
\end{description}

\begin{thebibliography}{99}

\bibitem{Brodsky}
	S. J. Brodsky and G. P. Lepage,
	\Journal{\PRL}{53}{545}{1979};
	Phys. Lett. {\bf 87B}, 359 (1979);
	G.P. Lepage and S.J. Brodsky,
	\Journal{\PRD}{22}{2157}{1980}.
\bibitem{Isgur}
	N. Isgur and C. H. Llewellyn Smith,
	\Journal{\PRL}{52}{1080}{1984};
	\Journal{\PLB}{217}{535}{1989};
	\Journal{\NP}{B317}{526}{1989}.
\bibitem{Radyushkin}
	A. V. Radyushkin,
	\Journal{\NP}{A523}{141c}{1991}.
\bibitem{Ji}
	C.-R. Ji, A. F. Sill, and R. M. Lombdar-Nelson,
	\Journal{\PRD}{36}{165}{1987};
	C.-R. Ji and F. Amiri, {\it ibid.}
	\Journal{}{42}{3764}{1990}.
\bibitem{Huang}
	T. Huang and Q. X. Sheng,
	\Journal{\ZPC}{50}{139}{1991}.
\bibitem{LiPion}
	H. N. Li and G. Sterman,
	\Journal{\NP}{B325}{129}{1992}.
\bibitem{LiProton}
	H. N. Li,
	\Journal{\PRD}{48}{4243}{1993}.
\bibitem{Jacob}
	R. Jacob and P. Kroll,
	\Journal{\PLB}{315}{463}{1993};
\bibitem{Caopion2}
	F.G. Cao and T. Huang, preprint hep-ph/9612233.	
\bibitem{CaoPion}
	F. G. Cao, T. Huang, and C. W. Luo,
	\Journal{\PRD}{52}{5358}{1995}.
\bibitem{CaoGamma}
	F. G. Cao, T.Huang, and B. Q. Ma,
	\Journal{\PRD}{53}{6582}{1996}.
\bibitem{Ji95}
	C.-R. Ji, A. Pang, and A. Szczepaniak,
	\Journal{\PRD}{52}{4038}{1995}.
\bibitem{Field}
	R. D. Field, R. Gupta, S. Otto, and L. Chang
	\Journal{\NP}{B186}{429}{1981}.
\bibitem{HMS}
T.~Huang, B.~Q.~Ma, and Q.~X.~Shen, Phys. Rev. D {\bf 49}, 1490 (1994).
\bibitem{MaZPA}
	B. Q. Ma,
	\Journal{\ZPA}{345}{321}{1993}.

\bibitem{MaJG}
	B.Q. Ma and T.Huang,
	J. Phys. G {\bf 21}, 765 (1995).

\bibitem{Wang}
	S. W. Wang and L. S. Kisslinger,
	\Journal{\PRD}{54}{5890}{1996}.

\bibitem{Dzie}
	Z. Dziembowski,
	\Journal{\PRD}{37}{778}{1987}.

\bibitem{Chung}
	P. L. Chung, F. Coester, and W. N. Polyzou,
	\Journal{\PLB}{205}{545}{1988}.


\bibitem{Kisslinger}
	L. S. Kisslinger and S. W. Wang,
	\Journal{\NP}{B399}{63}{1993};
	O. C. Jacob and L. S. Kisslinger,
	\Journal{\PLB}{243}{323}{1990};
	L. S. Kisslinger and S. W. Wang,
	Carnegie-Mellon preprint (1994), hep-ph/9403261.

\bibitem{Sch}
F. Schlumpf, Phys. Rev. {\bf D 50}, 6895 (1994).

\bibitem{crisis}
	B. Q. Ma, J. Phys. G {\bf 17}, L53 (1991);
	B. Q. Ma and Q. R. Zhang,
	\Journal{\ZPC}{58}{479}{1993};
	S. J. Brodsky and F. Schlumpf,
	\Journal{\PLB}{329}{111}{1994}.
\bibitem{Ma96}
	B. Q. Ma,
	\Journal{\PLB}{375}{320}{1996}.

\bibitem{LCF}
See, {e.g.}, J. B. Kogut and D. E. Soper, Phys. Rev. {\bf D 1}, 2901 (1970);
J. B. Bjorken, J. B. Kogut, and D. E. Soper, 
{\sl ibid.} {\bf 3}, 1328 (1971);   
S. J. Brodsky, R. Roskies, and R. Suaya, {\sl 
ibid.} {\bf 8}, 4574 (1973).
\bibitem{BrodskyPQCD}
S. J. Brodsky and G. P. Lepage, 
{\it Perturbative Quantum  Chromodynamics}, edited by
A. H. Mueller (Singapore, World Scientific, 1989), p. 93.
\bibitem{BHL}
S. J. Brodsky, T. Huang, and G. P. Lepage, in {\it Particles and
Fields-2}, Proceedings of the Banff Summer Institute, Banff, Alberta,
1981, edited by A. Z. Capri and A. N. Kamal (Plenum, New York,1983), p. 143;
G. P. Lepage, S. J. Brodsky, T. Huang, and P. B. Mackenize, {\it ibid.},
p. 83;
T. Huang, in {\it Proceedings of XXth International Conference on High Energy
Physics}, Madison, Wisconsin, 1980, edited by L. Durand and L. G. Pondrom,
AIP Con. Proc. No. 69 (AIP, New York, 1981), p. 1000.
\bibitem{Wigner}
E. Wigner, Ann. Math. {\bf 40}, 149 (1939).
\bibitem{Melosh}
H. J. Melosh, Phys. Rev. D {\bf 9}, 1095 (1974).

\bibitem{Bro96}
        S.J. Brodsky and B.Q. Ma,  
	\Journal{\PLB}{381}{317}{1996}.

\bibitem{Drell}
	S. D. Drell and T.M. Yan,
	\Journal{\PRL}{24}{181}{1970};
	G. West,
	\Journal{\PRL}{24}{1206}{1970}.

\bibitem{BL2}
        S.J. Brodsky and G.P. Lepage,   
	\Journal{\PRD}{24}{2848}{1981}.

\end{thebibliography}
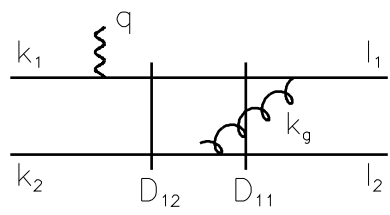
\end{document}